\documentclass[graybox]{svmult}
\usepackage{mathptmx}       
\usepackage{helvet}         
\usepackage{courier}        
\usepackage{type1cm}        
\usepackage{makeidx}         
\usepackage{graphicx}        
\usepackage{multicol}        
\usepackage[bottom]{footmisc}
\begin{document}

\title{Social Media Influence Operations}
\author{Raphael Meier}

\institute{{Raphael Meier} \at{Cyber-Defence Campus, armasuisse Science + Technology, Thun, Switzerland}, \email{raphael.meier@armasuisse.ch}}

\maketitle
\label{social_media_influence_operations}


\section{Introduction}
\label{sec:label1}
Affordable mobile devices, widely available internet connection, and social media platforms constitute modern Information Communication Technologies (ICTs). ICTs have fundamentally changed the way we communicate \cite{Herring2002}. In particular, social media platforms enable many-to-many communication without traditional gatekeeping mechanisms and theoretically little constraints on time and space. Through the use of language (and other means of communication), users of social media platforms can exchange information, engage in collective sense-making, and mobilize fellow users around a common interest. Online conversations gave rise to new social phenomena such as internet activism, crowdfunding, and open-source investigations. Machine learning algorithms, while increasingly embedded within modern societies, have so far been unable to effectively engage in online conversations. The introduction of prompt-based Large Language Models (LLMs) is changing that. According to Niklas Luhmann \cite{Baraldi2021}, language is the medium that structurally couples social systems (e.g. politics, education, etc.) with the psychological system of the individual human mind. Through the use of language, the human mind and social systems influence each other and co-evolve. Following Luhmann's theory, one can argue that we are now entering an era in which computer algorithms, in particular LLMs, are much stronger structurally coupled to the human mind and social systems than ever before\footnote{This line of argumentation is inspired by a recent commentary of the philosopher Hans-Goerg Moeller. Source: https://www.youtube.com/watch?v=9dNVmPepATM.}. Hence, LLMs have the potential to exert much stronger influence on individuals and social systems than previous computer algorithms did, which makes them an attractive tool for threat actors conducting influence operations online.

An influence operation conducted on social media can be seen as a concerted effort by an actor to interfere in an adversary's process of meaning-making through exploiting technical means provided by social media platforms \cite{Bergh2020}. Depending on doctrinal grounds of the actor, it can be regarded as an operation in cyberspace and/or the information space/environment, and it is typically performed covertly \cite{Stout2017}. In order to avoid the multitude of related terms (e.g. information operation, information warfare, etc.) and stick to an established definition, for the remainder of this article the concept of cyber-enabled social influence operations (CeSIOs) is being used \cite{Cordey2019}, which focuses on operations that utilize cyberspace '\textit{to shape public opinion and decision-making processes through the use of social bots, dark ads, memes and the spread of disinformation}.'

More recently, cyber operations have been recognized as a tool of subversion \cite{Maschmeyer2023}. In particular, it was shown that cyber operations are best suited to implement a slow-burning strategy of erosion of the adversary's strengths, which includes the erosion of public confidence for its government \cite{Shandler2022}. When used for malicious purposes, LLMs are subversive in their nature. The salience of LLMs enables an actor to use them covertly and to actively interfere in online communication. They considerably add to the deceptive properties of fake accounts on social media \cite{Mink2022}, so called sock puppet accounts, and as such exhibit the capabilities to subvert online discourse.

While there are already excellent overviews on the risks posed by LLMs (e.g. \cite{Weidinger2022}) and their general significance for influence operations \cite{Goldstein2023}, the purpose of this article is to present a concise summary of LLMs' salience and its potential impact on the instrumentation of sock puppet accounts---a core component of CeSIOs.

\section{LLMs' salience}
\label{sec:label2}
The misuse of LLMs to enhance influence operations, disinformation and propaganda was recognized early on (e.g. through generation of synthetic news articles \cite{Zellers2019}). Every release of a new, more capable LLM leads to a reiteration on potential misuses, which typically includes influence operations and related issues. For example, the introduction of GPT-3 \cite{Brown2020} sparked discussions on its potential to automatize the creation of disinformation \cite{Buchanan2021}, improve cost-effectiveness for threat actors \cite{Tamkin2021}, and generate toxic language \cite{Gehman2020}. In general, LLM-written text needs to satisfy three basic requirements for it to be useful to threat actors conducting CeSIOs: i) convey the intended message, ii) be persuasive, and iii) be very hard to distinguish from human written text (non-detectability).

An algorithmic hallmark of modern LLMs is their ability to engage in conversational fine-tuning (also called Reinforcement Learning with Human Feedback) \cite{Ouyang2022}. Earlier LLMs were plagued by the generation of unhelpful, inappropriate or outright toxic content, which was misaligned with the user's original intent. The ability to align machine output with user intent through fine-tuning based on human preferences is a crucial ingredient to generate messaging that reflects the threat actors intentions. Consequently, this increases not only the quality of the generated content but potentially also reduces the burden of post-editing and/or manual selection of LLM-written text through human operators of the threat actor.

There is an increasing amount of evidence demonstrating the persuasiveness of LLM-written text \cite{Burtell2023}. Bai et al. \cite{Bai2023} found that messages generated by GPT-3 were as persuasive as messages authored by humans in influencing study participants in their support for different policy issues (e.g. assault weapon ban, paid-parental leave). Furthermore, GPT-3 demonstrated the ability to produce propagandistic articles which were, with limited human curation, nearly as persuasive as articles stemming from state-sponsored influence campaigns \cite{Goldstein2023b}. The generation of persuasive messages through GPT-3 can further be augmented by taking into account the psychological profile of the intended target audience \cite{Matz2023}, rendering personalized persuasion at scale feasible. Jakesch et al. \cite{Jakesch2023} investigated a new, subtle type of influence, called latent persuasion, in which an opinionated LLM assisted study participants in expressing their own thoughts and ultimately shifting their opinions (i.e. aligning it with the opinion encoded in the LLM). Interestingly, LLMs themselves also exhibit a certain receptiveness to persuasive techniques (e.g. Illusory Truth Effect was demonstrated in GPT-3 \cite{Griffin2023}). 

Regarding non-detectability, it has been shown repeatedly that humans are unable to distinguish human-written from LLM-written text (e.g. \cite{Clark2021})  and rely on flawed cognitive heuristics while doing so \cite{Jakesch2023b}. It has also been shown that automatic detection of LLM-written text remains an open problem \cite{DaSilva2023}. A very recent analysis of a malicious botnet utilizing ChatGPT on Twitter reconfirmed that state-of-the-art LLM text detectors are currently unable to distinguish human-written from LLM-written text \cite{Yang2023}.

In the following section, the implications of LLMs' salience for the instrumentation of sock puppet accounts will be explored. Given the continuous evolution of both CeSIOs and LLMs, this endeavor is of course speculative.

\section{Potential impact}
\label{sec:label3}
Previous studies have shown that LLMs are most effective for covert influence when paired with human operators curating/editing their output and thus increasing quality of the output \cite{Buchanan2021,Goldstein2023b}. The employment of such a modus operandi for the instrumentation of sock puppet accounts by threat actors will likely yield two main effects in the near future: i) the amplification of old processes/tactics, and ii) increased operational security for sock puppet accounts.

Automated generation of text content with relatively high quality will enable sock puppet accounts to amplify the intensity of old processes/tactics such as e.g. astroturfing. By exploiting conversational fine-tuning, sock puppet accounts could be adapted towards engaging specific topics and/or target audiences with more tailored content. Hence, the burden of manual creation of text content by human operators could be reduced and these resources could be put to use elsewhere by the threat actor. The property of LLMs to engage in conversations increases their ability to dynamically respond to posts online (e.g. in comment section) and generate sophisticated responses in form of reviews/critics (e.g. in case of opinion summarization \cite{Bhaskar2022}). The latter is particularly concerning since second-order observations\footnote{The concept was introduced by Niklas Luhmann. We engage in second-order observation when we observe observations of others (e.g. when reading reviews of an interesting AirBnB space). In fact, this very article is based on second-order observation.} (e.g. academic peer review, restaurant reviews, influencer marketing, etc.) are used pervasively in modern societies to judge quality, and to construct and validate personal identity via the use of online profiles (see work on profilicity by Moeller et al. \cite{Moeller2021}). It is entirely conceivable that LLM-instrumented sock puppet accounts could be leveraged by threat actors to game any meaning-making process that is textual, based on second-order observations, and takes place on social media.

LLM-written text content will likely contain less language discrepancies, less copy-and-pasted text \cite{Goldstein2023}, and other idiosyncrasies that otherwise would jeopardize the operational security of the sock puppet account. Remaining issues (e.g. self-revealing messages) could be taken care of through automated filtering \cite{Yang2023} and human-assisted editing/curation. In addition, LLMs exhibit already basic capabilities for impersonation (e.g. taking on the role of an ornithologist) \cite{Salewski2023}. Consequently, attribution efforts based on textual content may become increasingly difficult or even infeasible.

Finally, LLM-instrumented sock puppet accounts or "counterfeit people"\footnote{A term recently coined by the philosopher Daniel Dennett for chatbots. Source: \url{https://www.theatlantic.com/technology/archive/2023/05/problem-counterfeit-people/674075/}} will potentially have a number of malicious psychological and societal consequences. A full review is beyond the scope of this article and priority is given to the most subversive of consequences: the liar's dividend \cite{Chesney2019}. Besides false information, synthetic content may also increase general skepticism towards objective truth, which will make it easier to discredit authentic content online as 'AI-generated'. Hence, a substantial increase of synthetic content online, propagated by sock puppet accounts, would normalize this behavior such that it could be used by threat actors to avoid accountability (liar's dividend). Moreover, first results point towards a decreased trustworthiness of online profiles when suspected to be AI-generated \cite{Jakesch2019}. This will likely yield an online environment with heightened uncertainty and greater difficulty in figuring out whom and what information to trust. Consequently, collective sense-making processes, especially in times of crisis (e.g. during Arab Spring in 2011 \cite{Kavanaugh2016}), which are taking place on social media could be subverted by CeSIOs using LLM-instrumented sock puppet accounts. Furthermore, well-informed citizens have long been regarded as a prerequisite for a functioning democracy \cite{Kuklinski2000}. Given that people increasingly inform themselves online, including via social media (50\% of US adults did it at least sometimes in 2022\footnote{Source: \url{https://www.pewresearch.org/journalism/fact-sheet/social-media-and-news-fact-sheet}}), CeSIOs married with the capabilities of advanced LLMs will make it even harder for the individual to navigate the information space on social media. In a most pessimistic outlook, this could contribute to a slow erosion of public confidence in democratic (e.g. judiciary system, free press) and epistemic institutions (e.g. universities) (cf. \cite{Goldstein2023}). While pondering on the potential impact, it is, however, important to keep in mind that the potency of CeSIOs is still being contested, with some studies suggesting strong limitations (e.g. \cite{Maschmeyer2023b}), while others suggest tangible real-world consequences (e.g. \cite{Starbird2023}).

\section{Mitigation}
\label{sec:label4}
An excellent overview on potential mitigations for influence operations using LLMs can be found in the publication by Goldstein et al. \cite{Goldstein2023}. When focusing on LLM-instrumented sock puppet accounts, three specific mitigations should be highlighted: i) limiting available infrastructure for threat actors, ii) characterization of behavioral patterns, and iii) introduction of watermarked LLM-written content.

First, it is important to realize that while a threat actor may more or less easily create a sock puppet account and propagate LLM-written content, it is by no means guaranteed that this content will reach a sufficient mass of the target audience. In order to achieve that, the threat actor needs to build a network of sock puppet accounts with sufficient reach and credibility. Making it harder for threat actors to create and cultivate sock puppet accounts would drastically mitigate the impact of CeSIOs (e.g. by requiring proof-of-personhood \cite{Goldstein2023}, or by limiting access to relevant training data for LLM fine-tuning). Second, while LLM-written text content may be hard to detect, behavior of LLM-instrumented sock puppets may still offer an avenue to identify anomalies. Recent work showed that sock puppets using ChatGPT belonging to the same suspected Twitter botnet were identifiable through commonalities in their inauthentic, coordinated behavior \cite{Yang2023}. Based on novel detection methods, social media platforms could then take appropriate steps in case of suspicious activity (e.g. suspension of account). Third, work on new ways to watermark LLM-written content should be intensified \cite{Kirchenbauer2023}. Secure and widely-adopted watermarking of LLM-content would prevent covert use and likely reduce effectiveness of LLM-instrumented sock puppet accounts for CeSIOs.

However, mitigations aimed at reducing the presence of sock puppet accounts on social media are inevitably in conflict with business incentives of social media platforms, which traditionally are most interested in an ever-growing user base and activity. Hence, new norms and laws on AI, social media and truthfulness \cite{Evans2021}, which deem LLM-instrumented sock puppets illicit, are much needed to cause a rethinking of current business incentives and the potential social harm they are causing.

At last every one of us is responsible for not putting more oil into the fire when using social media \cite{Lima2022}. We are responsible when we share insights into our lives online, opening us up to potential manipulation, when we amplify the reach of a false news article on which we have read not more than the headline (and that conveniently validates our misconceptions), and when we fuel heated online arguments that alienate a group of users. We need to better recognize such moments and resist the urge to post. Moving forward, awareness on personal responsibility of social media use and media literacy should be created among the general public to increase society's resilience against future CeSIOs.

\bibliographystyle{unsrt}
\bibliography{social_media_influence_operations_ref}


\end{document}